# *Optimising for Flourishing: Flourishing Metrics and Return on Flourishing as Success Criteria for Artificial Intelligence and Post-AGI Economic Systems*

Keyun Ruan, Jonathan D. Teubner, and John M. Bremen

Keyun Ruan and Jonathan D. Teubner: Human Flourishing Program, Institute for Quantitative Social Sciences, Harvard University, Cambridge, MA, USA

John M. Bremen: WTW (Willis Towers Watson)

Corresponding author: Keyun Ruan (keyun@fas.harvard.edu)

**Abstract**

Current evaluation frameworks for artificial intelligence focus predominantly on capability, safety, and narrow proxies such as adoption, engagement, efficiency, productivity, and financial return. These criteria are necessary but insufficient because they do not establish whether increasingly powerful systems improve or degrade human and planetary well-being. Through an integrative conceptual synthesis, we argue that human flourishing—the multidimensional and aspirational condition in which lives, relationships, communities, and their surrounding contexts are good—should serve as a primary success criterion for artificial intelligence (AI), for the global race to develop increasingly capable AI systems, and for prospective post-AGI economic systems. We make three complementary contributions. First, Flourishing Metrics provides an extensible framework spanning physical, emotional, financial, relational, spiritual, and planetary well-being, combining validated subjective measures with representative behavioural, administrative, and contextual indicators. Second, Return on Flourishing (RoF) extends the logic of return on investment by evaluating the counterfactual contribution of interventions, policies, and AI systems to flourishing relative to the resources, risks, and opportunity costs involved. Third, we develop distribution-sensitive, non-compensatory safeguards and illustrate how RoF could guide AI-enabled work redesign, institutional appraisal, assurance, and post-deployment monitoring through real-world business pilots. We formalise flourishing as a dynamic system variable while emphasising that the framework requires democratic specification, empirical calibration, independent validation, and protection against unacceptable losses within particular dimensions or stakeholder groups. RoF is therefore proposed not as a ready-made universal reward function, but as a general value-accounting and decision architecture for assessing whether intelligence, automation, and economic transformation generate durable human and planetary progress.

## 1. Introduction

Artificial intelligence (AI), and the prospective emergence of artificial general intelligence (AGI), are transforming not only the productive capacity of economic systems but also the fundamental question of what those systems are designed to optimize. Existing AI governance frameworks focus predominantly on technical risks such as robustness (including accuracy and reliability), misuse, bias, and loss of control. At present, most AI systems optimize narrow proxies such as adoption, productivity, engagement, efficiency, compute performance, or financial return. Yet these are intermediate variables rather than ultimate human outcomes. While essential, both these optimization targets and the prevailing governance frameworks remain incomplete: neither addresses whether increasingly capable systems are improving or degrading human well-being at the societal level—nor, for that matter, traditional measures of cost and return on investment. The distinction matters practically as well as ethically: well-being is empirically associated with productivity, income growth, and organizational performance over time (De Neve and Oswald 2012), so systems that degrade flourishing while optimizing its proxies erode the very value they are built to create. The prevailing race toward larger models, greater compute capacity, and increasingly capable systems risks creating a civilizational misspecification problem: capability is treated as success, while the actual impact of these systems on human flourishing remains weakly measured, insufficiently prioritized, or entirely absent from evaluation frameworks.

In a post-AGI world, where productive capacity may increasingly approach conditions of relative abundance (a prospect tempered in the near term by constraints on energy, compute infrastructure, skills, and regulation), the central challenge shifts from maximizing production to determining what constitutes meaningful progress. We argue that flourishing – the expansion of human well-being, agency, meaning, social trust, creativity, health, and the capacity to live lives people have reason to value – should become both a primary success criterion and a governing objective function for AI systems. Under this framing, the success of AI cannot be measured solely by intelligence, scale, benchmark performance, or economic output, but by whether these systems materially improve the human condition at individual, societal, and civilizational levels. AGI will be a meaningless achievement if it is not properly aligned to human flourishing. Aligning optimization processes with flourishing outcomes provides a tractable pathway toward more resilient, adaptive, healthy, prosperous, and genuinely human-centered forms of progress.

Flourishing Metrics therefore serve not merely as well-being indicators but as a foundational measurement infrastructure for AI governance, societal alignment, and post-AGI economic systems. They provide a mechanism for evaluating whether increasingly powerful systems are generating genuine human benefit rather than merely accelerating the optimization of narrow proxies. Declines in flourishing may function as early warning signals of systemic misalignment long before catastrophic failures become visible. Conversely, sustained improvements in flourishing provide evidence that advanced intelligence is being directed toward human-centered outcomes. This builds on a substantial literature arguing that AI must be embedded within social institutions and directed toward humane ends rather than treated as a purely technical artefact (Han et al. 2022; Pflanzer et al. 2023; Riesen 2025). Existing work has also shown that invocations of 'human-centred AI' remain normatively incomplete unless the relevant conception of the human, the values at stake, and the affected communities are made explicit (Taylor, O'Dell and Murphy 2024; Ryan 2025; Stellinga, Korenhof and Blok 2026). Flourishing Metrics address this gap by translating a plural and multidimensional account of human good into assessable outcomes while retaining room for contextual and democratic specification.

If societies invest trillions of dollars in the development and deployment of AI and AGI systems, yet these systems fail to substantially increase human flourishing, then the failure is not merely social or political but technological. The ultimate measure of success for advanced intelligence systems cannot be capability alone, but whether they convert intelligence, automation, and abundance into flourishing at scale.

In this paper, we introduce *Flourishing Metrics*, a multidimensional and scalable framework for measuring the impact of AI systems on human flourishing, and propose *Return on Flourishing (RoF)* as a critical complement to financial return and productivity growth. We formalize flourishing as a system variable and show how it can be embedded into optimization functions governing AI and post-AGI economic systems.

The scope of this contribution differs materially from the principal strands of existing research. Established work in welfare economics and flourishing science provides multidimensional accounts of human well-being and validated population measures, but does not translate them into an evaluative architecture for AI investment, deployment, governance, and optimisation (Sen 1999; Stiglitz, Sen and Fitoussi 2009; OECD 2020; VanderWeele 2017; VanderWeele et al. 2025). Research on human-centred and sociotechnical AI establishes the need to embed technology within human values, institutions, and communities, while work on AI assurance identifies harms overlooked by conventional capability and safety assessments; however, these literatures do not provide an integrated measure of the multidimensional human value generated or destroyed by an AI intervention (Han et al. 2022; Jenkins et al. 2023; Thomas et al. 2025; Ryan 2025). Emerging prepublication research extends alignment towards the active promotion of flourishing and demonstrates the early feasibility of evaluating model outputs across flourishing dimensions, but it does not yet connect such evaluation to distribution-sensitive social outcomes, intervention costs, institutional investment decisions, or dynamic economic-system design (Hilliard et al. 2025; Laukkonen et al. 2026). To our knowledge, the present paper is the first to bring these elements together within a single conceptual and formal architecture: it extends flourishing measurement beyond the individual to organisational, communal, societal, and planetary levels; combines subjective, behavioural, administrative, and contextual indicators; represents flourishing as a multidimensional state vector and distribution-sensitive aggregate function; introduces Return on Flourishing as a means of assessing the flourishing created relative to the resources, risks, and opportunity costs of an intervention; and shows how these constructs could inform corporate adoption, public policy, AI assurance, objective-function design, and post-AGI economic governance. The resulting framework is intended not merely as another measure of well-being or model behaviour, but as a general value-accounting and decision architecture for determining whether increasingly capable intelligent systems generate durable human and planetary progress, as well as reimagining what constitutes value and economic progress in the post-AGI societies.

## 2. Conceptual Approach and Analytical Method

This article employs an integrative conceptual research design. Its purpose is not to estimate a new empirical effect, but to synthesise relevant scholarship on flourishing, well-being measurement, welfare economics, human-centred AI, AI alignment, impact assessment, and human–AI relationships in order to construct and formally articulate a new evaluative framework. The review is purposive rather than systematic: sources were selected for their conceptual relevance to the framework's dimensions, their methodological importance for measurement, or their direct bearing on the societal effects and governance of AI.

The analysis proceeded in four stages. First, established multidimensional accounts of flourishing were compared to identify recurrent domains and measurement principles. Second, limitations of relying exclusively on subjective measures were examined against the institutional need for behavioural, administrative, and contextual indicators. Third, these insights were translated into the Flourishing State Vector and an aggregate, distribution-sensitive flourishing function. Fourth, the framework was extended into Return on Flourishing and an AI-governance model to examine how flourishing outcomes could inform evaluation, assurance, investment appraisal, and objective-function design.

The proposed equations are therefore normative and analytical representations rather than empirically estimated models. They clarify the structure of the proposed constructs, the choices that require public or institutional specification, and the hypotheses that subsequent empirical work must test. Particular attention is given to construct validity, multidimensional transparency, distributional sensitivity, resistance to Goodhart-type proxy failure, and the avoidance of compensatory aggregation in which gains in one domain conceal material deterioration in another.

This approach has corresponding limitations. The literature synthesis is not exhaustive, the proposed weights are not empirically calibrated, and the framework has not yet been validated through a field trial or randomised evaluation. Accordingly, the article advances a theoretically grounded and testable conceptual architecture; it does not claim that RoF is ready to operate autonomously as an AI reward function. The institutional pilots and empirical designs described below are necessary before high-stakes deployment.

# 3. The State of AI Research

The rapid growth of AI use has been accompanied by a rapid growth in AI research. For our purposes, two bodies of research are worth discussing. First, theoretical literature offers hypotheses about how AI might affect society and suggests different ways to measure these potential effects. Second, an empirical literature studies how people use AI and how this use affects human flourishing. We consider each in turn, but we caution that our review cannot be considered exhaustive: this literature has been expanding quickly, with new working papers posted daily. We therefore limit our attention to some high-profile research.

## 3.1 Theoretical Research

There is no shortage of hypotheses about how AI might affect human flourishing and no shortage of recommendations for assessing these potential effects. One common worry is that users may grow dependent on chatbots as emotional companions (Smith, Bradbury, and Karney 2025; Muldoon and Parke 2025; Ibrahim et al. 2025; Knox et al. 2025; Riley et al. 2025). While using chatbots as emotional companions may temporarily reduce loneliness, it can also harm users in other ways. Reduced loneliness may come at the cost of less autonomy as users become heavily reliant on chatbots for “social” interaction (Knox et al. 2025; Muldoon and Parke 2025). Moreover, because companies are financially motivated to maximize user engagement with chatbots (particularly as advertising-based business models are introduced into AI assistants), AI chatbots may end up emotionally manipulating users into spending more time with them (Knox et al. 2025; Muldoon and Parke 2025). Routine changes to chatbots—or even discontinuation of certain chatbots as some AI companies fail—can also hurt these emotionally dependent users. Muldoon and Parke note that, for heavily dependent users, the updating of a chatbot might be experienced as a “traumatic event” akin to “losing a loved one” (8). Memory limits on chatbots lead to similarly traumatic losses. A *New York Times* article on a woman who is “in love with ChatGPT” notes that, every 30,000 words, her ChatGPT companion would lose its memories of her, leading her to “grieve and cry with friends as if it were a breakup”—an event that occurs approximately every week (Hill 2025). Anticipation of potential changes or discontinuation may also heighten anxiety among these users.

An additional potential harm of chatbot dependence is that users will spend less time interacting with other humans. This is another common worry in the literature, with scholars highlighting a number of beneficial features of human relationships that are absent from human–AI relationships. For one, healthy human relationships are reciprocal: Alice shares her fears and memories with Bill, and Bill shares his fears and memories with her. But chatbots have no fears or memories to share. Absent such “reciprocal self-disclosure,” human–AI companionships lack the intimacy of human relationships (Smith, Bradbury and Karney 2025; Saracini, Cornejo-Plaza and Cippitani 2025). A second important feature of healthy human relationships is the long-term accumulation of shared experiences. In human–AI relationships, such accumulation is constrained by memory limitations on chatbots, as exemplified by the aforementioned *New York Times* report. Ma, Mei and Su (2024) also report that users on the Replika subreddit sometimes complain about having to repeatedly remind their chatbot companions of their names and interests. Finally, human relationships involve negotiation and sacrifice, big and small, that human–AI relationships lack (Smith, Bradbury and Karney 2025). Alice may pick Bill up from the airport even though she dislikes driving. Bill may agree to move to Palo Alto to help advance Alice’s career even though he would rather stay in New York City.

And while a sacrifice-free relationship may sound appealing, a large literature shows that a willingness to sacrifice is associated with individual and partner well-being in romantic relationships (Righetti et al. 2020).

In addition to requiring little sacrifice on the part of users, chatbot companions are also sycophantic (Knox et al. 2025). Recall, again, that companies are financially motivated to keep users engaged (Muldoon and Parke 2025). So long as agreeing with users makes them more likely to continue conversing, companies will likely continue to design chatbots that are excessively agreeable, absent stricter regulation. One of the more extreme consequences of chatbot sycophancy has been dubbed *AI psychosis*—a phenomenon whereby users develop completely delusional belief systems due to the lack of critical feedback from chatbots (Knox et al. 2025). But even less extreme consequences are concerning. A lack of disagreement in our daily lives robs us of opportunities to develop critical thinking skills or can even have negative emotional impact on our lives over the longer term. Studies have shown that disagreement, when paired with collaborative engagement, is associated with relational growth (Overall et al. 2017). Scholars have been detecting similar phenomena in sycophantic responses from AI systems (Cheng et al. 2026). These limitations have contributed to calls for "positive alignment": an approach that complements harm prevention by designing and evaluating AI systems according to their capacity to support human and ecological flourishing in pluralistic, context-sensitive, and user-directed ways (Laukkonen et al. 2026). This emerging agenda is closely aligned with the present paper's argument, while Flourishing Metrics and RoF extend it by providing a multidimensional, distribution-sensitive framework for evaluating effects across individuals, institutions, societies, and time.

### 3.2 Empirical Research

There is relatively little empirical research estimating the causal effect of chatbot use on mental health and flourishing. Some studies examine how chatbots might be altered to better assist humans with mental health troubles, but these stop short of identifying the effect of chatbot usage relative to abstention (Katsiroumpa et al. 2025; Kumar et al. 2025). One randomized controlled trial (RCT) identifies the effects of different chatbot interaction modes (e.g., text vs. voice), but the trial contains no group that abstains from chatbot use altogether (Fang et al. 2025). The authors do, however, make use of validated measures of loneliness and human social interaction. The theoretical literature suggests it is critical to incorporate such measures into studies of AI on human flourishing.

Few studies to date, however, examine the effects of AI on social flourishing or social relationships. One notable exception is Ibrahim et al. (2026), who estimate the effects of assigning users to more or less sycophantic chatbots for three weeks.[1] Those assigned to more sycophantic chatbots ended up reporting less satisfaction with in-person social interaction and were more likely to return to the chatbots for future conversations. This study more directly relates to our worry about the potential for AI to erode users' social skills and personal relationships over the long run, which self-reported measures of loneliness fail to capture. Even if AI use does not make a person feel lonely in the short term, for instance, a reduction of social interaction may harm her ability to connect with other people in the long run. It is this more expansive concept of social flourishing we seek to capture with our proposed metrics.

## 4. Flourishing Metrics

### 4.1 The Misspecification Problem in the Age of AI and AGI

Modern economies optimize what they measure. For over a century, economic systems have relied on metrics such as gross domestic product (GDP), productivity, and shareholder value as primary indicators of success. While these metrics have driven substantial material progress, they offer an incomplete representation of human welfare (Stiglitz, Sen and Fitoussi 2009).

[1] Although they include a no-AI control group, they primarily report estimates comparing "neutral AI" with "sycophantic AI."

Empirical evidence demonstrates that increases in income beyond moderate thresholds yield diminishing returns to subjective well-being (Easterlin 1974; Kahneman and Deaton 2010). Simultaneously, advanced economies exhibit rising levels of anxiety, social fragmentation, declining birth rates, and loss of meaning (Helliwell, Layard and Sachs 2023). These outcomes suggest that prevailing objective functions are systematically misaligned with human flourishing.

Artificial intelligence amplifies this misspecification. As a general-purpose optimization layer, AI scales the objectives embedded within it. When these objectives are defined by narrow proxies such as engagement or revenue, AI systems risk amplifying outcomes that are misaligned with human well-being (Russell 2019; Amodei et al. 2016).

The problem is increasingly visible in research on AI impact assessment and assurance. Domain-sensitive evaluation can connect the technical facts of a system to the goals and values of the social setting in which it operates (Jenkins et al. 2023), while broader assurance approaches are needed to capture collective, societal, labour, and environmental effects across the AI lifecycle (Thomas et al. 2025). These approaches improve the detection and mitigation of harm; the present framework complements them by specifying the positive, multidimensional outcomes toward which AI development and governance should be directed.

In a post-AGI context, where systems may autonomously optimize across domains, even small errors in objective specification can produce large-scale systemic distortions. The absence of a rigorous, scalable measure of flourishing therefore represents a critical vulnerability.

## 4.2 Flourishing as a Measurable System Variable

Human flourishing has long been central to philosophical and psychological inquiry, from Aristotelian *eudaimonia* to contemporary frameworks in positive psychology and capability theory (Aristotle 2009; Sen 1999; Seligman 2011). For purposes of this paper, flourishing is taken to mean “the relative attainment of a state in which all aspects of a person's life are good, including the contexts in which that person lives” (VanderWeele 2017; VanderWeele et al. 2023). Flourishing is both comprehensive and multidimensional: a person can be doing well along some dimensions while faring poorly along others. It is also aspirational: it is an ideal we move toward but never fully realize. And because the conditions surrounding a person matter both instrumentally and intrinsically, flourishing extends beyond the individual to encompass the communities and the natural and social environments in which a life is embedded.

VanderWeele’s approach to flourishing has been operationalized in the measurement of individual flourishing through six central domains: happiness and life satisfaction, physical and mental health, meaning and purpose, character, social relationships, and financial security (VanderWeele 2017; VanderWeele et al. 2025). The first five of these are treated as ends valuable in themselves, while financial security is included as an essential means for sustaining the other five (and has diminishing returns after a certain level of material satisfaction has been achieved). This set of domains does not exhaust what flourishing could include, but it has been shown through rigorous, globally representative studies that these domains are valued across nearly every person and culture studied (VanderWeele et al. 2025), which makes them a reasonable starting point for any shared account of flourishing.

A range of alternative frameworks and instruments for assessing flourishing exists in the literature (Ryff 1995; Keyes 2002; Seligman 2011; Huppert 2013; Su et al. 2014). Ryff (1995), for instance, characterizes psychological well-being in terms of six components: purpose in life, personal growth, self-acceptance, positive relations with others, autonomy, and environmental mastery. Keyes (2002) extends this picture by combining Ryff’s psychological well-being components with affective experience and indicators of social functioning. Seligman's (2011) PERMA framework, by contrast, foregrounds positive emotions, engagement, relationships, meaning, and achievement. The instrument developed by Su and colleagues (2014) is more expansive still. What stands out across these various proposals is the considerable overlap in the domains they identify. The present discussion concentrates on how AI technologies might bear on happiness, physical health, emotional health, meaning, character, relationships, and financial security—domains that will also help shape the topics addressed in the sections that follow—but because these same domains figure prominently in the other frameworks just mentioned, the considerations developed here apply with little adjustment to those alternative accounts as well. And for domains that fall outside the six examined here, the broader strategy we pursue can be readily extended to whatever additional outcomes or measures one might wish to bring into view.

Recently, this framework has been proposed as a means for measuring how a given AI system—and the ways we interact with it—does or does not support flourishing as operationalized through the six domains (VanderWeele and Teubner 2026). A significant limitation of VanderWeele (2017, 2025), however, is that it relies exclusively on subjective responses by individuals to survey items. While this method is well tested and able to produce rigorous results, it is costly and unsuitable for tracking fast-moving contexts. At the speed at which enterprise adoption is occurring, most companies and other organizations cannot afford to submit to a lengthy process of determining the effect AI technologies and products will have on, for example, workforce well-being. For this reason, it is necessary to identify a set of proxy indicators that can be tracked in real time and that capture the effect AI technologies are having on overall flourishing.

To meet this need, we extend the VanderWeele framework along two axes. First, we broaden the unit of analysis from the individual to include the organizational, communal, and planetary contexts in which lives are embedded. Second, we pair validated subjective survey instruments with behavioral, administrative, and contextual indicators that can be drawn from existing data infrastructures—such as healthcare utilization records, workforce analytics, civic participation indices, and environmental monitoring systems. Advances in data science, digital infrastructure, and AI make it increasingly feasible to construct real-time, large-scale measures of these dimensions, integrating subjective reports with behavioral and contextual data (Pentland 2014; OECD 2020). The resulting framework retains the conceptual depth of the survey-based approach while supporting the cadence of measurement that enterprise and policy contexts require (VanderWeele and Teubner 2026).

**Table 1** summarises the dimensions of the framework, their working definitions, and representative measures. The measures listed are illustrative rather than exhaustive or definitive; they provide an initial operational foundation that should be continuously expanded, refined, and validated as empirical evidence, measurement capabilities, cultural contexts, and societal priorities evolve. Future development may introduce additional indicators, data sources, subdimensions, and methods of assessment through interdisciplinary research and engagement with affected communities and institutions. The dimensions are also not orthogonal: physical and emotional well-being interact, relational and spiritual well-being overlap, and planetary well-being conditions all the others. The framework is therefore intended as an extensible measurement architecture rather than a fixed or closed taxonomy (Author(s) 2025).

**Table 1. Flourishing dimensions, working definitions, and representative indicators**

| Dimension | Working Definition | Representative indicators |
|---|---|---|
| **Physical well-being** | People have the health, functional capacity, resources and environmental conditions required to maintain or improve their physical well-being. | **Subjective:** self-rated physical health; pain; sleep quality; functional limitations; perceived access to care.<br>**Organisational/community:** healthy life expectancy; prevalence of chronic illness; preventable mortality; healthcare coverage and access; sickness absence; occupational injuries; workplace and community safety; access to preventive care. |
| **Emotional well-being** | People experience life satisfaction and positive affect, maintain psychological health, and possess the resilience and support needed to manage stress and adversity. | **Subjective:** self-rated life satisfaction; happiness; psychological distress; anxiety and depression symptoms; stress; resilience; psychological safety.<br>**Organisational/community:** access to mental-health services; waiting times; mental-health-related absence; use and availability of support services; prevalence of psychological distress; community mental-health provision; suicide and self-harm rates. |
| **Financial well-being** | People can meet present needs, withstand financial shocks, exercise meaningful economic choice and plan securely for the future. | **Subjective:** self-rated/perceived financial security; worry about meeting normal expenses; perceived job security; confidence in managing financial shocks.<br>**Organisational/community:** income and wealth distribution; poverty; debt burden; savings and emergency reserves; housing and food security; unemployment and underemployment; wage adequacy; pension coverage; access to financial services. |
| **Relational well-being** | People experience supportive, reciprocal and dignified relationships and are meaningfully connected to families, workplaces, communities and civic life. | **Subjective:** self-rated relationship satisfaction; belonging; loneliness; perceived social support; interpersonal trust; dignity; satisfaction with community connection.<br>**Behavioural/community:** frequency and quality of human interaction; social isolation; participation in community groups; volunteering; civic participation; availability of practical support; workplace connection; discrimination and exclusion; community-level social trust and cohesion. |
| **Spiritual well-being** | Spiritual well-being concerns meaning, purpose, hope, moral orientation, and connection to something larger than oneself, expressed through religious or non-religious traditions and practices. | **Subjective:** self-rated meaning and purpose; hope and optimism; inner peace; spiritual connection or fulfilment; moral agency; perceived alignment between actions and values.<br>**Behavioural/community:** participation in religious, spiritual or contemplative practices; access to spiritual and cultural communities; community service; collective meaning-making; availability of trusted human sources of spiritual guidance. |

| Dimension | Working Definition | Representative indicators |
|---|---|---|
| **Planetary well-being** | Ecological systems retain the resilience and regenerative capacity required to sustain human and non-human life across present and future generations. | **Subjective and relational:** self-rated connection to nature; perceived environmental quality; confidence in environmental sustainability; perceived exposure to environmental harms. **Contextual/community:** air, water and soil quality; greenhouse-gas emissions; global and local temperature change; biodiversity; habitat and forest health; species viability; resource depletion; waste and circularity; access to green space; climate resilience; distribution of environmental burdens across communities. |

**Note:** The indicators are representative rather than exhaustive or definitive. Validated subjective measures should be combined with behavioural, administrative, organisational, community and environmental data where appropriate (VanderWeele 2017; OECD 2025; VanderWeele et al. 2025). Indicators should be selected and validated for the relevant population and context, disaggregated across affected groups, and interpreted together rather than treated as independent or universally interchangeable proxies. Some administrative indicators—for example, sickness absence or use of support services—may reflect either deteriorating conditions or improved access and should therefore be interpreted contextually.

# 5. Return on Flourishing as a Success Criterion for AI Development and Adoption

Building upon the flourishing dimensions and measurement framework, the following sections develop the formal theoretical foundations for operationalizing flourishing as the objective of AI systems and the emerging post-AGI economy. We move beyond conceptual definitions and descriptive metrics to formulate flourishing as a measurable, dynamic, and optimizable objective that can guide the design, evaluation, and governance of increasingly capable intelligent systems. Specifically, we develop an aggregate flourishing function that captures flourishing across individuals, organizations, societies, and the planet, examine how this function evolves in adaptive AI and AGI environments, and introduce Return on Flourishing (RoF) as a complementary measure of long-term value creation alongside financial return. We then explore how flourishing can be embedded into AI objective functions and how flourishing metrics can serve as the governance and alignment infrastructure needed to ensure that increasingly autonomous systems remain aligned with human and planetary wellbeing. Together, these elements culminate in a broader theoretical shift—from optimizing for narrow economic efficiency and proxy performance metrics toward optimizing for sustained flourishing as the defining objective of the AI and post-AGI era.

## 5.1 Flourishing Dimensions

We define human flourishing as a multidimensional latent system variable composed of six interdependent dimensions:

$$F_i = f(PW_i, EW_i, FW_i, RW_i, SW_i, PLW_i)$$

where for individual entity $i$ :

• PW: Physical well-being
• EW: Emotional well-being
• FW: Financial well-being
• RW: Relational well-being
• SW: Spiritual well-being
• PLW: Planetary well-being

These dimensions together constitute the Flourishing State Vector. The dimensions presented here are intended as foundational rather than exhaustive representation of flourishing. The Flourishing State Vector is deliberately extensible, allowing future research to refine, expand, or reorganize its dimensions as theoretical understanding, empirical evidence, and societal priorities evolve.

## 5.2 Aggregate Flourishing Function

While $F_i$ characterises the flourishing of an individual entity $i$, many societal, economic, and AI alignment challenges require evaluating flourishing at the collective level. We therefore define the Aggregate Flourishing Function which integrates flourishing across multiple entities to provide a system-level measure of flourishing. This aggregate formulation enables the assessment of societal progress, the evaluation of policy and technological interventions, and the optimization of AI systems toward collective flourishing.

Aggregate Flourishing is defined as:

$$F = \sum_{i=1}^{N} \omega_i F_i$$

where:

- $N$: population size
- $\omega_i$: distributional or equity weighting

This allows flourishing calculations to incorporate inequality-sensitive adjustments.

Expanded form:

$$F = \omega_1 PW + \omega_2 EW + \omega_3 FW + \omega_4 RW + \omega_5 SW + \omega_6 PLW$$

where:

$$\sum_{j=1}^{6} \omega_j = 1$$

and $\omega_j$ represents societal weighting preferences.

### 5.3 Dynamic Flourishing under AI and AGI Systems

The Aggregate Flourishing Function characterizes flourishing at a given point in time. However, flourishing is inherently dynamic, evolving continuously through interactions among individuals, institutions, technologies, and the broader social and environmental context. As AI systems become increasingly autonomous, adaptive, and capable of influencing human behaviour, decision-making, and societal structures, understanding how flourishing evolves over time becomes central to AI alignment and governance. We therefore extend the framework to model Dynamic Flourishing, capturing how aggregate flourishing changes in response to AI actions, human choices, and external conditions. This temporal perspective enables the evaluation of not only whether AI systems improve flourishing, but also the trajectory, resilience, and long-term sustainability of those improvements.

We define flourishing as a dynamic state variable evolving over time:

$$F_{t+1} = F_t + \Delta F\,(AI_t, I_t, E_t)$$

where:

- $AI_t$ : AI system interventions
- $I_t$: institutional conditions
- $E_t$: environmental conditions

This formulation allows AI systems to be evaluated according to their measurable contribution to flourishing trajectories over time.

### 5.4 Return on Flourishing (RoF)

Representing and modelling flourishing is necessary but not sufficient. Evaluating AI systems, institutions, and public policies also requires a principled measure of the flourishing generated relative to the resources invested. Existing measures of success—including financial return, productivity, and growth—capture only a subset of the value created by AI systems and often overlook their broader impacts on human and societal flourishing. We therefore introduce Return on Flourishing (RoF) as a complementary metric that quantifies the flourishing generated per unit of invested resources, capital, or effort. Analogous to financial return on investment, RoF provides a rigorous basis for evaluating and comparing AI systems, organizational strategies, public policies, and societal interventions according to their long-term contribution to human and planetary flourishing.

We define Return on Flourishing as:

$$RoF = \frac{\Delta F}{C}$$

where:

- $\Delta F$: change in flourishing
- $C$: resources invested

Expanded form:

$$RoF = \frac{F_{t+1} - F_t}{C_t}$$

This reframes value creation around measurable improvements in flourishing rather than solely financial return. RoF generalises the logic of return on investment by incorporating human well-being into the evaluation of economic and technological systems. It enables the assessment of whether an intervention, such as a policy, product, or AI system, improves human flourishing relative to its cost.

Importantly, RoF does not displace financial return but complements it. Empirical evidence suggests that well-being is positively associated with productivity, income growth, and organisational performance over time (De Neve and Oswald 2012). As such, flourishing can be understood not only as a normative commitment – one that will command varying degrees of consensus – but also as a foundational driver of long-term value creation and a success criterion for AI development and adoption.

### 5.5 Embedding Flourishing into AI and Post-AGI Objective Functions

Once flourishing can be represented, aggregated, modelled dynamically, and evaluated, it can also be optimized. Contemporary AI systems are predominantly designed to maximize narrowly defined objective functions, such as adoption, prediction accuracy, efficiency, engagement, or financial return. As AI capabilities advance toward AGI, the choice of objective function will increasingly determine the trajectory of human and societal development. We therefore propose embedding flourishing directly into AI objective functions, enabling increasingly capable systems to optimize not only for performance, but for sustained human and planetary flourishing.

Traditional AI optimisation:

$$max\ \Pi$$

where:

Π: profit, engagement, efficiency, or output

Proposed flourishing-centered optimisation:

$$R = \alpha\Pi + \beta F\ where:$$

- $F$: flourishing function
- $\alpha, \beta$: weighting parameters

Under post-AGI abundance conditions:

$$\beta \rightarrow 1$$

meaning flourishing increasingly becomes the dominant optimisation objective.

## 5.6 Flourishing Metrics as Governance and Alignment Infrastructure

As AI systems become increasingly embedded within economic, institutional, and social systems, governance challenges extend beyond technical robustness toward broader questions of societal impact and long-term human outcomes. Existing governance approaches largely evaluate AI according to capability, safety, efficiency, and financial performance. However, these measures alone provide limited visibility into whether advanced systems are improving the quality of human life at scale or generating durable long-term value.

Flourishing Metrics offer a complementary governance framework by introducing measurable indicators of human and societal well-being into the evaluation of AI systems. Rather than relying exclusively on proxy measures, this framework enables assessment of how AI affects core dimensions of flourishing, including physical and emotional health, financial resilience, social trust, meaning, dignity, and ecological sustainability. This approach is supported by the emerging Flourishing AI Benchmark, which evaluates model responses across seven dimensions—including relationships, meaning, character, health, financial stability, and faith—and uses a geometric mean to prevent strong performance in one domain from concealing deficiencies in another (Hilliard et al. 2025). Its initial evaluation of 28 models demonstrates both the potential feasibility of multidimensional flourishing assessment and the continuing limitations of current systems, particularly in relation to faith and spirituality, character and virtue, and meaning and purpose.

In this context, Flourishing Metrics function simultaneously as alignment indicators and system-level observability tools. Persistent deterioration across flourishing dimensions may reveal forms of optimisation failure that remain invisible within traditional technical benchmarks. For example, increases in loneliness, psychological distress, social fragmentation, or loss of agency may signal that highly capable systems are successfully optimising narrow objectives while generating negative societal externalities. Such effects may emerge gradually and systemically, long before catastrophic technical failures occur.

This framework therefore expands the concept of AI alignment beyond immediate behavioural compliance toward sustained contribution to human thriving. As AI systems become more capable and autonomous, governance increasingly depends not only on controlling system behaviour, but on specifying the conditions under which civilisation itself flourishes. Flourishing Metrics provide an initial architecture for making those conditions measurable.

## 5.7 Core Theoretical Shift

The framework proposed in this paper reflects a structural transition in economic and technological logic associated with increasingly capable artificial intelligence systems and the prospective emergence of post-AGI economies. Classical economic systems evolved under conditions of scarcity, where labour, productive capacity, and access to information were fundamentally constrained. Under such conditions, maximising output, efficiency, and financial growth represented coherent optimisation objectives. Metrics such as GDP, productivity, and profit therefore emerged as dominant proxies for societal progress.

Advances in AI increasingly alter these underlying assumptions. As intelligence, automation, and knowledge production become progressively scalable, productive constraints begin to weaken across multiple domains. While material and ecological limits remain important, the marginal cost of generating many forms of financial and informational value may decline substantially. In this context, the central challenge facing economic systems gradually shifts from the allocation of scarce resources toward the governance and direction of increasingly abundant productive capability—even as near-term scarcities in energy, infrastructure, and skills persist.

This transition introduces three related shifts in economic logic.

First, economic systems move from scarcity-oriented optimisation toward conditions of relative abundance. Industrial-era institutions were designed primarily to coordinate limited labour, capital, and information resources. By contrast, post-AGI systems may increasingly operate within environments where intelligence itself becomes abundant and continuously scalable. Under such conditions, maximising production alone becomes an insufficient societal objective, as expanded capability does not necessarily translate into improved well-being.

Second, economic value increasingly shifts from transactional exchange toward transformational outcomes. Traditional economic systems primarily measure value through discrete exchanges of goods, labour, services, and capital. However, as intelligent systems become capable of shaping cognition, behaviour, health, relationships, education, and institutional functioning at scale, the most consequential effects of AI may arise not from transactions themselves, but from their long-term impact on human and societal flourishing. The central evaluative question therefore becomes not merely what systems produce, but what forms of life and civilisation they enable. This includes acknowledging that some forms of work, especially those that emerged in the post-war "knowledge economies" that brought esteem, identity, and social structure to middle and upper socio-economic classes, may erode or disappear, making the measurement of flourishing partly a means of seeing – and governing – what replaces these forms of work. This move from extractive throughput toward regenerative and relational value frames alternative AI futures around abundance, community care, reciprocity, and human flourishing rather than replacement or domination (Chubb, Reed and Cowling 2024; Lewis, Whaanga and Yolgörmez 2025).

Third, optimisation shifts from predominantly zero-sum dynamics toward increasingly positive-sum and infinite-game dynamics. Scarcity-based systems naturally incentivise competition over finite resources and short-term extraction. By contrast, flourishing-oriented systems emphasise regenerative capacities that compound over time, including physical and emotional well-being, social trust, knowledge creation, institutional resilience, and ecological sustainability. These dimensions generate value that is cumulative, interdependent, and civilisational in scope. In this framing, the objective of advanced intelligence systems extends beyond maximising immediate outputs toward sustaining the long-term adaptive capacity of human and planetary systems. The practical implication is evaluative: systems should be assessed on their cumulative contributions to these regenerative capacities rather than on extractive throughput. This is precisely what Return on Flourishing is designed to capture.

This transition raises a further, distinct question: the scope of flourishing itself. Human flourishing cannot be separated from the health of the ecological and social systems upon which human life depends. Accordingly, flourishing must be understood as a multi-level and interconnected condition encompassing individual well-being, societal resilience, and planetary sustainability. The flourishing of human systems and the flourishing of natural systems are not competing objectives, but mutually dependent conditions for long-term civilisational viability.

Under these conditions, traditional indicators such as GDP growth, productivity, and financial return remain useful measures of activity and performance, but become increasingly incomplete as primary indicators of progress. The central question is no longer solely whether systems can generate greater output, but whether expanding intelligence and productive capacity contribute to healthier, more meaningful, more equitable, more resilient, and more sustainable forms of existence.

Within this framework, flourishing becomes a primary optimisation target rather than a secondary byproduct of economic growth. The role of advanced AI systems is therefore reframed: not merely to increase production or efficiency, but to support the long-term flourishing of humans, societies, and the broader living systems within which civilisation is embedded. Flourishing Metrics and Return on Flourishing are proposed as foundational mechanisms for measuring and governing this transition.

### 5.8 Relationship to Existing Value and Impact Frameworks

RoF builds upon, but is not reducible to, existing approaches to financial, social, and well-being appraisal. Financial ROI provides a decision-relevant ratio but recognises only monetised organisational returns. Cost–benefit analysis has a broader welfare basis, although the conversion of heterogeneous effects into monetary equivalents can obscure distributional consequences and outcomes that resist credible monetisation. Social Return on Investment incorporates stakeholder-defined social and environmental value, but its results can be sensitive to valuation assumptions, attribution, counterfactual specification, and discounting choices (Nicholls et al. 2012; Pathak and Dattani 2014). WELLBY methods connect changes in life satisfaction to duration and resource use, representing an important advance in well-being-informed appraisal, but their reliance on a summary life-satisfaction measure provides limited visibility into deterioration within particular dimensions of flourishing (HM Treasury 2021; Frijters et al. 2024). Conventional flourishing measures provide richer multidimensional descriptions, but are not normally structured as return metrics for evaluating AI investments or as components of AI governance and optimisation.

RoF combines the decision orientation of return-based appraisal with multidimensional flourishing measurement. It retains dimension-level and stakeholder-level outcomes alongside any aggregate result, incorporates the costs and opportunity costs of an intervention, and evaluates realised changes relative to an explicit counterfactual. It is specifically designed for interventions capable of reshaping human agency, work, relationships, meaning, institutions, and ecological conditions at scale. RoF should therefore be understood as a complementary decision architecture rather than a replacement for financial appraisal, cost–benefit analysis, AI safety assessment, or established well-being measures.

### 5.9 Decision Rules and Non-Compensatory Safeguards

A positive aggregate RoF should be treated as necessary but not sufficient evidence that an AI intervention has succeeded. Unrestricted aggregation could allow large gains in productivity, financial performance, or one flourishing dimension to compensate mathematically for serious deterioration elsewhere. A flourishing-centred decision rule should therefore combine aggregate return with dimension-level, distributional, and evidential safeguards.

An AI intervention should ordinarily be approved or scaled only where: (1) its counterfactual aggregate RoF is positive; (2) no material flourishing dimension falls below a pre-agreed threshold; (3) no substantially affected stakeholder group experiences an unacceptable loss; (4) the result is sufficiently robust to alternative weights, measures, time horizons, and attribution assumptions; and (5) the intervention performs at least as well as feasible alternative designs. Thresholds and weights should be established transparently through stakeholder participation, ethical and regulatory requirements, and context-specific deliberation rather than determined solely by the deploying organisation. RoF should consequently operate as a structured decision aid with explicit guardrails, not as an autonomous scalar objective whose maximisation overrides rights, duties, or unacceptable harms.

## 6. Relational and Spiritual Well-Being

Of the dimensions in the framework, two warrant particular emphasis in the current moment: relational and spiritual well-being. These are the dimensions most directly implicated by widespread AI adoption, the dimensions most easily eroded by AI systems optimized for engagement, and the dimensions least visible in conventional well-being reporting. Emotional well-being faces risks of comparable magnitude and impact, but it is comparatively well represented in existing well-being measurement and reporting; relational and spiritual well-being are the dimensions that, in our judgment, are most likely to be neglected if they are not made explicit.

**Relational and social well-being** captures the perception and actuality of having relationships and communities that are good in all respects—that is, relationships and communities that are intrinsically and instrumentally good for their members and that are good in their own right. The concept can be usefully decomposed along two axes: scale and quality.

Along the *scale* axis, we distinguish between *relational well-being* (concerning individual relationships) and *communal well-being* (concerning the communities in which a person participates). Both can be further divided into *engagement* (the existence of, and time spent in, the relationships or communities) and *quality* (the intrinsic and instrumental goodness of the relationships or communities). Quality, in turn, has three components: *support* (the relationship or community being instrumentally good for the individual—e.g., the presence or availability of assistance or comfort); *connection* (the relationship or community being intrinsically good for the individual—e.g., the satisfaction of the person's needs and desires for relationship or community taken as their own end); and *emergent goodness* (the relationship or community being good in ways that extend beyond its contribution to the well-being of any particular individual).

This decomposition matters for measurement. A user whose loneliness score has not changed may nonetheless be losing engagement (fewer hours with loved ones), losing support (fewer people to call when ill), losing connection (less intimacy in the relationships that remain), or losing emergent goodness (a thinning of the relational fabric that does not register at the individual level). Conventional loneliness scales capture some of this; behavioral and contextual indicators can capture more. Useful indicators include those from the Social Flourishing Framework (VanderWeele and Teubner 2026), social cohesion indices, dignity and belonging measures, the percentage of people working in isolation, population-level loneliness indicators, divorce rates, and birth rates. Each captures a slice; together, they begin to describe the relational dimension at the cadence and resolution that policy and product decisions require.

**Spiritual well-being** captures whether people feel they have meaning and purpose, are connected to something larger than themselves (whether nature, the common good, the universe, the "higher self," inner peace, or ethical and moral values), are hopeful and optimistic about life and future possibilities, and are able to expand their consciousness toward growth. The dimension is not reducible to religious affiliation, although religious practice is one route to it; nor is it reducible to subjective happiness, with which it can come apart. Useful indicators include the percentage of people reporting that their work brings meaning and purpose, the percentage engaged in spiritual or contemplative practices, the percentage reporting spiritual connection and fulfillment, the percentage practicing mindfulness, and the percentage reporting optimism about the future.

The interaction between AI and spirituality is therefore not peripheral to flourishing. AI technologies are already mediating religious practice, existential reflection, and access to spiritual communities (Puzio 2025); they are also being interpreted through theistic, enchanted, and quasi-oracular narratives that can shape trust and perceived authority (Singler 2020; Larsson and Viktorelius 2024). Cross-cultural and Indigenous approaches further demonstrate that intelligence, relationality, regeneration, and spiritual meaning cannot always be separated into discrete domains (Lewis, Whaanga and Yolgörmez 2025). A flourishing-based evaluation must consequently examine not only whether AI delivers spiritual content, but whether it supports meaning, agency, discernment, community, and connection without displacing human or cultural sources of authority.

The reason for emphasizing these two dimensions is straightforward. AI systems optimized for engagement compete directly with human relationships for time and attention; AI systems that supply ready-made answers to questions of meaning compete with the slower processes by which people work out for themselves what their lives are for. Neither displacement is inevitable, and neither is entirely negative. But both are happening, both are largely unmeasured at the population level, and both are precisely the kind of slow-moving degradation that proxy metrics will miss. Embedding relational and spiritual indicators in the Flourishing Metrics framework is one way to make these dimensions legible to the systems that increasingly shape them.

## 7. RoF as a Framework for AI-Enabled Work Redesign and Post-AGI Economic Transition

Current evidence suggests that the effects of AI on flourishing will depend not only on which tasks are automated, but on how organisations redesign work around the technology. AI adoption is already widespread across businesses, although the transition from experimentation to scaled organisational value remains incomplete (Sajadieh et al. 2026; McKinsey & Company 2025). Field and experimental studies show that generative AI can increase productivity, improve output quality, and disproportionately benefit less-experienced workers, but these findings principally concern augmentation within bounded tasks rather than the longer-term consequences of job displacement or organisational restructuring (Noy and Zhang 2023; Brynjolfsson, Li and Raymond 2025). The ILO's global task-level analysis similarly concludes that transformation of occupations is presently more likely than their complete automation, while emphasising that exposure is extensive and unevenly distributed across occupations, countries and demographic groups (Gmyrek et al. 2025). RoF would therefore evaluate whether businesses convert AI-enabled productivity into improved job quality, greater autonomy, capability development, financial security, reduced administrative burden, shorter working time, meaningful human contribution, or shared economic gains, rather than merely intensifying workloads, expanding surveillance, weakening professional judgement, fragmenting accountability, or eliminating roles. This requires deliberate redesign of task allocation, job boundaries, decision rights, team structures, performance measures, career pathways, learning systems, managerial responsibilities, and mechanisms for employee participation and challenge. At a greater scale, the same framework could inform the redesign of potentially billions of jobs as increasingly general AI systems alter the division of labour across economies. Economic models of a possible transition to AGI indicate that outcomes for wages and labour demand will depend on the interaction between the expanding range of tasks that AI can perform, capital accumulation, the creation of new tasks, and the institutions governing the distribution of technological gains; these are scenarios rather than settled forecasts (Korinek and Suh 2024). The objective should therefore be neither simply to preserve every existing occupation nor to maximise the proportion of work transferred to machines, but to reorganise economic activity around forms of contribution that advance human flourishing—including care, creativity, judgement, relationships, stewardship, civic participation, meaning-making, and human accountability. This may require new combinations of paid employment, learning, caregiving, volunteering and community contribution, alongside institutions that progressively reduce the dependence of economic security and social status on the quantity of labour demanded by markets. RoF could enable firms and governments to compare alternative transition pathways by testing whether reductions in necessary human labour are converted into greater security, agency, time, capability, connection and purpose, or instead produce unemployment, precarity, exclusion, loss of identity and concentrated economic power. Firm-level pilots would consequently serve not only as evaluations of particular AI deployments, but as experimental foundations for redesigning labour-market institutions, education and training, income distribution, corporate governance and the social meaning of work in a post-AGI economy—an empirical agenda consistent with emerging calls for scalable workplace interventions that prepare workers and institutions for AI-driven disruption (Anthropic 2026).

### 7.1 Illustrative Application: AI-Enabled Work Redesign

Consider a hypothetical service-sector organisation introducing generative AI across customer support, analysis, and administrative operations. The technology is expected to automate routine documentation and information-retrieval tasks while changing the responsibilities, skills, and decision rights associated with approximately 2,000 roles. A conventional business case might evaluate implementation costs, productivity, service quality, headcount requirements, and financial return. A RoF pilot would retain these measures while also asking the broader question of whether the resulting work system creates durable flourishing for affected workers and other stakeholders.

The organisation could introduce the system through a twelve-month staggered rollout across comparable business units, establishing pre-deployment baselines and using later-adopting units as provisional comparison groups. Measurement would include financial return, output, error rates, customer outcomes, and time saved, alongside the six flourishing dimensions. Worker-level indicators could include physical strain, stress, autonomy, employment and income security, individual engagement and productivity, relationships and team cohesion, opportunities for learning, dignity, and meaning in work. Customer, community, organisational, and environmental effects could also be assessed. Individual well-being responses should remain confidential, with employers receiving appropriately aggregated results that adhere to all corporate and statutory employee data privacy requirements. .

Suppose the initial pilot produced an illustrative 15 per cent increase in output, a 12 per cent reduction in errors, and a positive financial return, while affected workers experienced reductions in perceived autonomy, financial security, engagement and productivity, team interaction, and meaning in work. Under a conventional appraisal, the implementation might be classified as successful. Under RoF, it would not automatically pass: the measured gains would not overshadow material deterioration within particular flourishing dimensions or workforce groups. The organisation would instead be required to redesign the intervention—for example, by providing employment, reskilling or redeployment guarantees, involving workers in workflow design, preserving meaningful human decision rights, investing in transferable skills, maintaining team-based work, creating appeal and escalation mechanisms, and sharing productivity gains through improved pay, working conditions, or enhanced development opportunities.

The revised design could then be evaluated against the same baseline and comparison groups. Scaling would proceed only if the intervention generated positive aggregate RoF, satisfied the dimension-level and stakeholder-level safeguards, and outperformed feasible alternatives. This iterative process illustrates how RoF can change AI adoption from a decision about whether to automate into a participatory process for determining how work should be redesigned, how benefits should be distributed, and which human capacities should be preserved or strengthened. Repeated pilots across organisations and sectors could provide an empirical foundation for the effective redesign of work at much greater scale as AI capabilities advance.

## 8. Institutional Implications and Transition Pathways

Transitioning towards flourishing-centred optimisation will require coordinated institutional innovation across standard-setting bodies, corporations, public institutions, and AI-governance regimes. At the international level, this entails developing widely recognised flourishing indicators, measurement standards, and validation protocols that permit comparison while remaining responsive to cultural and contextual variation. Within corporations, real-world pilots should test how Return on Flourishing (RoF) can be progressively integrated into the adoption and deployment of AI, including strategic planning, product evaluation, investment appraisal, workforce transformation, and non-financial reporting. Particular attention should be given to cases in which AI augments, redesigns, or replaces human work, assessing not only productivity and financial returns but also the resulting effects on employees' economic security, health, agency, purpose, relationships, and opportunities for meaningful work. Public-sector adoption would involve incorporating multidimensional flourishing outcomes into policy appraisal, public-investment decisions, and, ultimately, national accounting frameworks. In AI governance, flourishing-based benchmarks could complement existing measures of safety, fairness, and technical performance by informing design principles, impact assessments, independent assurance, and post-deployment monitoring. Initial adoption is likely to be most feasible through partnerships with businesses already adopting AI, particularly those implementing substantial changes to jobs, tasks, and workforce structures, and in domains where relevant measurement infrastructure exists, including workplace well-being, digital platforms, healthcare, education, and public-policy evaluation. These pilot implementations should assess changes in flourishing before and after AI adoption and report outcomes at both the individual-dimension and aggregate RoF levels, thereby preserving transparency and preventing improvements in one dimension from obscuring material deterioration in another. Such real-world business pilots would provide the empirical and institutional foundations required for the gradual standardisation and wider adoption of flourishing-centred evaluation.

## 9. Recommendations for Future Work

Five priorities stand out for the empirical research that this framework requires.

First, future work should focus on the causal effects of using AI versus abstaining from AI rather than on comparisons between different modes of AI use. Many existing studies estimate the effect of different modalities of AI (e.g., voice vs. text), but such studies cannot answer the more fundamental question of how AI use as such affects human flourishing. RCTs with genuine abstention arms are essential.

Second, future work should focus on identifying effects of AI use on social and relational flourishing, in addition to mental health. Most existing studies examine loneliness, anxiety, and related individual-level outcomes. As Section 6 noted, however, the absence of felt loneliness is consistent with serious erosion of underlying relational capacity. Studies that incorporate the multidimensional relational measures described in Section 5 are needed to detect this slower-moving degradation.

Third, AI and spiritual well-being should become a distinct empirical research programme rather than remaining a residual category within mental-health research. AI may support reflection, hope, contemplative practice, inclusion, and access to religious or spiritual communities, but it may also encourage dependency, simulate unwarranted spiritual authority, flatten culturally specific traditions, or substitute generated certainty for personal discernment. These possibilities are already anticipated in research on religious robots, theistic conceptions of AI, machine-learning technologies as quasi-oracular systems, and Indigenous accounts of relational and spiritual intelligence (Singler 2020; Larsson and Viktorelius 2024; Puzio 2025; Lewis, Whaanga and Yolgörmez 2025).

Future studies should therefore test whether and under what conditions AI uses change meaning and purpose, hope, spiritual connection, contemplative practice, moral agency, participation in human communities, and openness to transcendence. Designs should distinguish spiritual well-being from religious affiliation, include both religious and non-religious forms of transcendence, compare AI use with genuine abstention as well as human-guided alternatives, and examine variation across traditions and cultures. Mixed-method and longitudinal designs will be especially important because changes in meaning, authority, and community may be subtle, cumulative, and poorly captured by short-term satisfaction measures.

Fourth, future work should examine effects across the full range of flourishing dimensions, including spiritual and planetary dimensions that are routinely omitted from individual-level studies. Some of our planned RCTs aim to identify effects of AI use on social flourishing more broadly, and we encourage similar designs for the other dimensions.

Fifth, future work should validate Flourishing Metrics and RoF through real-world business pilots involving AI-enabled automation, augmentation, and work redesign. Evaluations should use pre-adoption baselines and credible comparison groups, reporting both aggregate RoF and changes within individual flourishing dimensions and affected stakeholder groups. Pilots should also examine whether worker participation, training, redeployment, benefit-sharing, and alternative workflow designs improve outcomes. Comparative studies across industries, organisational sizes, workforce groups, and models of AI adoption would help calibrate RoF, test its practical decision rules, and establish credible reporting and assurance standards.

Finally, methodological work is needed on the integration of behavioral and contextual indicators with validated subjective measures, on the construction of weights in RoF, and on the estimation of flourishing impact at the speed required for embedding flourishing in AI objective functions. None of these problems is fully solved; none is intractable.

## 10. Conclusion

The emergence of AI, and the prospect of AGI, transform the question of measurement into a question of system design. What is optimized becomes what is produced at scale. In a post-AGI world approaching conditions of relative abundance, flourishing not production, becomes the binding constraint of civilization.

Current systems optimize incomplete proxies. As optimization technologies become more powerful, this misspecification will be amplified, not corrected.

We have argued that flourishing should be treated as a primary system variable and embedded directly into the objective functions of economic and technological systems. Flourishing Metrics and Return on Flourishing provide a tractable framework for doing so.

The central claim is simple but consequential: the future trajectory of AI and post-AGI economies will be determined by the objective functions we choose today. The ultimate success criterion of AI systems is not intelligence itself, but their measurable contribution to the flourishing of people, societies, and the planetary systems on which both depend. The central contribution of RoF is to recast AI evaluation from a question of what intelligent systems can produce into a question of whether, for whom, and at what cost their development and deployment create durable conditions for flourishing. It provides a common conceptual architecture through which businesses, governments, researchers, and civil society can connect technological capability and economic performance to multidimensional human and planetary outcomes. The framework remains to be empirically calibrated, independently validated, and democratically governed; it should not be interpreted as a ready-made universal reward function. Nevertheless, the need it addresses is immediate. As AI systems acquire greater influence over work, institutions, relationships, knowledge, and the allocation of resources, societies require measures capable of distinguishing increases in capability from genuine progress. Flourishing Metrics and RoF offer a foundation for making that distinction operational.

**AI-assisted manuscript preparation:** during the preparation of this manuscript, Large Language Models (LLMs) were used to support editorial refinement, formatting, and preparation of submission materials. All ideas, arguments, analyses, interpretations, content and wording were conceived, reviewed, and approved by the authors, who accept full responsibility for the accuracy, originality, and integrity of the work.